# A Solution Method for the Reynolds-Averaged Navier-Stokes Equation

T.-W. Lee

*Mechanical and Aerospace Engineering, SEMTE, Arizona State University, Tempe, AZ, 85287*

**Abstract- Using the Lagrangian transport of momentum, the Reynolds stress can be expressed in terms of basic turbulence parameters. DNS data at higher Reynolds numbers (Re$_\tau$= 1000 and 5200) have been used to again validate this theory, where it is the Lagrangian momentum balance between u'$^2$ and pressure fluctuation forces that determine the Reynolds stress at these conditions. This approach can be used to obtain key parameters such as the von Karman constant, inner layer thickness and the Reynolds stress itself. This expression for the Reynolds stress can be combined with RANS for an iterative solution for turbulent flows in channels. A viable solution for turbulence channel flow, iteratively obtained using this Lagrangian formalism, is presented.**

T.-W. Lee
Mechanical and Aerospace Engineering, SEMTE
Arizona State University
Tempe, AZ 85287-6106
Email: attwl@asu.edu

## INTRODUCTION

Determination of the Reynolds stress in terms of root turbulence parameters has profound implications in fluid physics and engineering, as it determines the mean flow structure through turbulence (fluctuating) momentum transport.  As is well known, turbulence models attempt to express the Reynolds stress in terms of known (or computable) parameters so that artificial closure is achieved.  The procedure is to expand the Reynolds stress or the turbulent viscosity, in terms of higher-order terms, e.g. turbulence kinetic energy, dissipation and third- and fourth-order correlations in the case of Reynolds stress models.  In turn, these higher-order terms need to be sequentially modelled [1-5].

Recently, we showed that a reverse is possible by considering the transport equation for turbulence momentum in a Lagrangian coordinate frame, i.e in a control volume moving with the local mean velocity [6]. For simple (canonical) geometries, this leads to an explicit expression for the Reynolds stress, u'v', albeit in terms of the main diagonal component, $u'^2$. Validation of this expression has been presented for canonical flow geometries, where the results agreed well with DNS and experimental data for channel, zero pressure-gradient flat plate and jet flows [6, 7]. Here, we show that this approach also yields the von Karman constants that are in close agreement with published values. In addition, we present an interative method where the expression for the Reynolds stress [6] makes it possible to compute the flow structure for turbulence channel flows, without resorting to modeling.

To review the current approach, we start by presenting an alternative perspective for the mechanics of the turbulence momentum transport, for the channel flow as an example. Experimental and computational observations show that $<u'^2>$ profiles have a peak near the wall, then gradually approach the centerline value. Attempts to find the right combination of U, $<u'^2>$, and other parameters for the Reynolds stress produced an entire library of turbulence models, at times with good results, as reviewable in the literature [1-5]. Figure 1(a) shows a very basic comparison of the Reynolds stress and $<u'^2>$ profiles, where $<u'^2>$ has been multiplied by a negative number to look at the two parameters side by side. The data are from DNS results for turbulent channel flows at $Re_\tau$ = 650 and 300 [8]. We can see that there are some hints that these profiles exhibit similarities in that both the Reynolds stress and $<u'^2>$ profiles exhibit a sharp peak near the wall. The peaks are slightly removed from one another, and the Reynolds stress goes to zero at the centerline while $<u'^2>$ reaches a finite value in the DNS data in Figure 1(a). However, to satisfy the centerline boundary conditions, the slopes of both of these profiles become zero at the centerline.

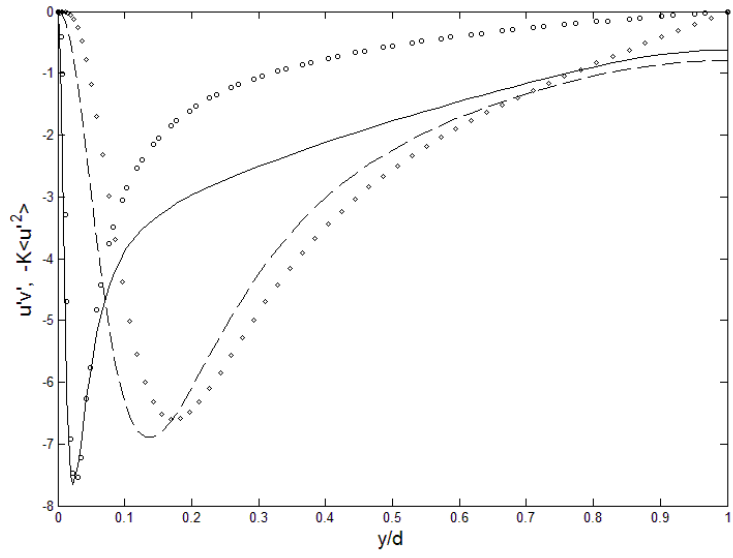

(a)

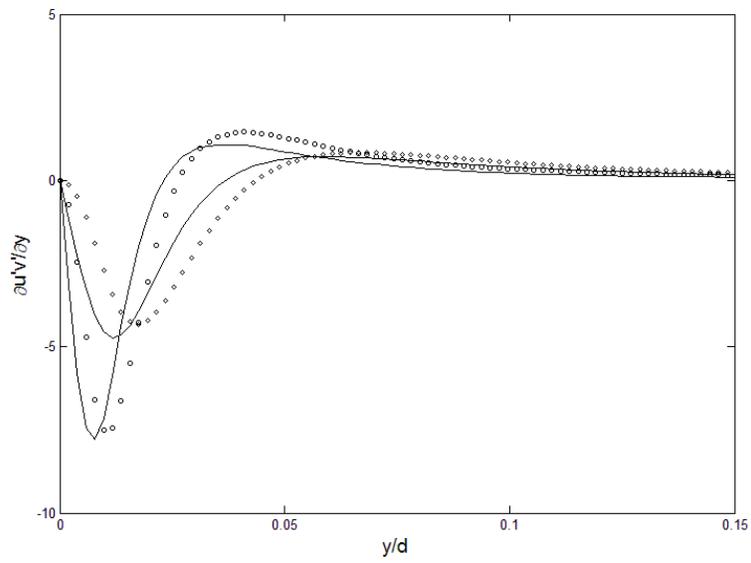

(b)

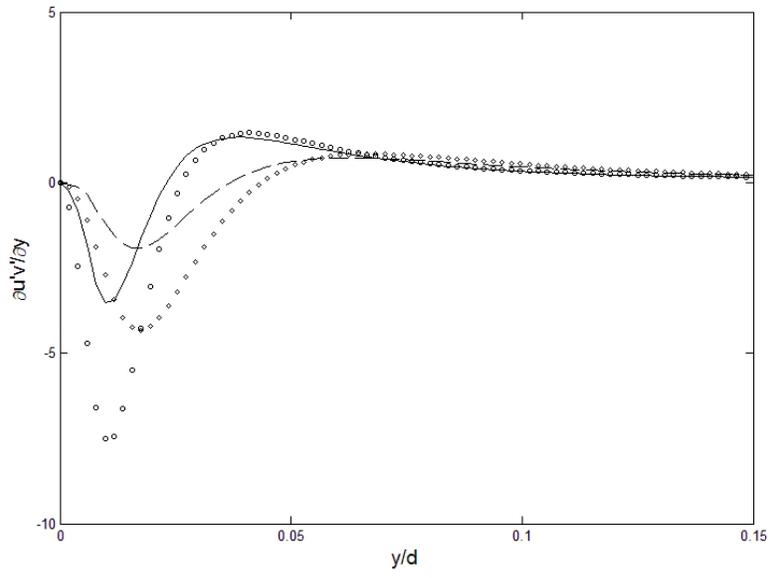

(c)

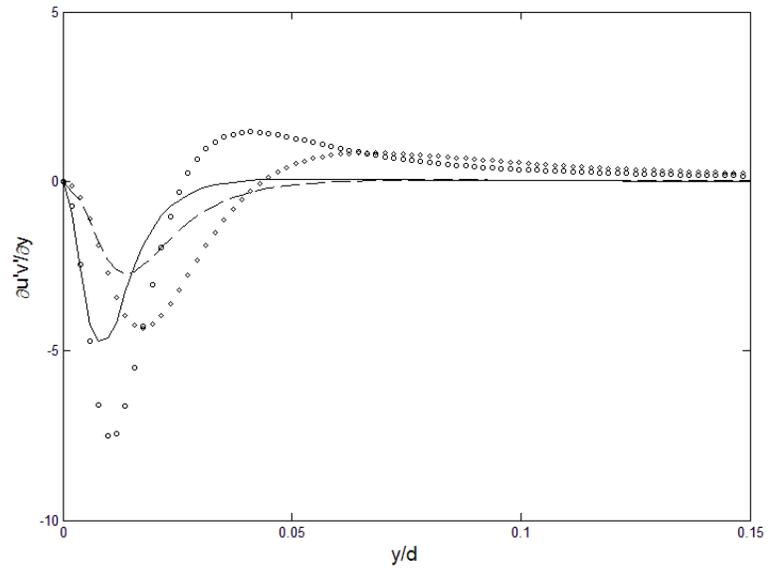

(d)

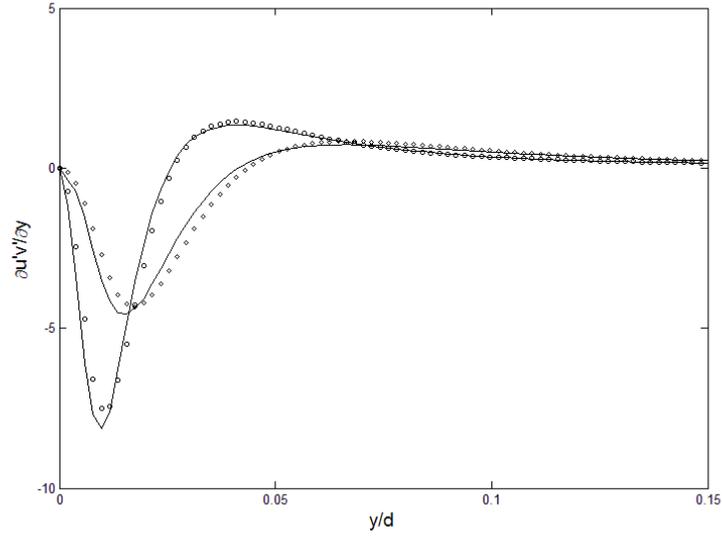

(e)

Fig. 1(a)-(e). Examination of the relationship between the Reynolds stress and <u'²> in various forms, for $Re_\tau$ = 650 (circle symbols) and 400 (diamond): (a) <u'v'> (symbols) and –K<u'²> (lines), K = 0.015; (b) $\dfrac{\partial <u'v'>}{\partial y}$ and $\dfrac{\partial <u'^2>}{\partial y}$; (c) $\dfrac{\partial <u'v'>}{\partial y}$ and $-C_1 U \dfrac{\partial <u'^2>}{\partial y}$; (d) $\dfrac{\partial <u'v'>}{\partial y}$ and $\nu_m \dfrac{\partial^2 u_{rms}'}{\partial y^2}$; and (e) $\dfrac{\partial <u'v'>}{\partial y}$ and $-C_1 U \dfrac{\partial <u'^2>}{\partial y} + \nu_m \dfrac{\partial^2 u_{rms}'}{\partial y^2}$. Symbols are from DNS data for <u'v'> [8], while lines represent various terms computed using <u'²> and U, also from the same DNS data set.

We can further observe the structural features in Figure 1(b) where the *gradients* of the Reynolds stress and <u'²>, in Figure 1(b). The slopes of these profiles become zero at the centerline, and even though the location for the minimum in the gradient is not quite aligned the overall shape of the profiles is quite similar. We can examine if such similarity is just a matter of coincidence or somehow related to the momentum transport by making a slight adjustment to $\dfrac{\partial <u'^2>}{\partial y}$. We multiply $\dfrac{\partial <u'^2>}{\partial y}$ by $C_1 U(y)$, where U is the mean velocity, and this is again compared with $\dfrac{\partial <u'v'>}{\partial y}$ in Figure 1(c). Multiplying by $C_1 U(y)$ magically aligns the peaks, and also the shape of the profiles become quite self-similar. As shown later, this is due to the transport of <u'²> by the local mean velocity U(y). To anticipate the mathematical formulation in the next section, we can also compare the gradient of the Reynolds stress with a viscous term, $\nu_m \dfrac{\partial^2 u_{rms}{}'}{\partial y^2}$, where u'$_{rms}$ is taken as the square root of <u'$^2$>. This is plotted in Figure 1(d), and again compared with $\dfrac{\partial <u'v'>}{\partial y}$. Since the <u'$^2$> profiles reach a nearly constant slope in Figure 1(a), we can deduce that the second derivative would go to zero away from the wall, as verified in Figure 1(d). Moreover, they have pronounced negative dips near the wall, in the so-called viscous sub-layer. Adding the two profiles in Figure 1(c) and (d) results in a very close agreement with the Reynolds stress gradient, as shown in Figure 1(e). The match is almost exact. From this analysis, it is seen that the contributions to the Reynolds stress are two-fold: from both <u'$^2$> and the viscous shear stress due to mean turbulent fluctuations near the wall, and only the <u'$^2$> effect is significant away from the wall. In channel flows, the sequence of Figures 1(a)-(e) shows that the Reynolds stress can be prescribed by U and <u'$^2$>, and no other variables or complex permutations of turbulence parameters are necessary. Thus, in spite of its simplicity Eq. 1 contains the correct dynamics that relates the Reynolds stress with two key turbulence parameters, the mean velocity and u'$^2$.

Mechanistically, the Lagrangian transport theory (or model) originates from consideration of a control volume moving at the local mean velocity [6, 7]. In this coordinate frame, the momentum consists of the fluctuating components: for example, for an observer moving with the mean flow velocity, the mean flow speed is not felt, but only the fluctuating components, in either x- and/or y-direction. Within this coordinate frame and logic, we obtain the following expression for the Reynolds stress gradient, after using a differential transform, $\dfrac{\partial}{\partial x} = C_1 U \dfrac{\partial}{\partial y}$, to account for the displacement effect in boundary layers [6]. Alternatively, we can substituting a transform in Navier-Stokes equations: U+u' $\rightarrow$ u', and V+v' $\rightarrow$ v'. Under this Galilean transform, we obtain

$$\frac{d(u'v')}{dy} = -C_1 U \left[ \frac{d(u'^2)}{dy} - \frac{1}{\rho} \frac{d|P|}{dy} \right] + \frac{1}{v} \frac{d^2 u'}{dy^2} \qquad (1)$$

In Eq. 1, all the terms are Reynolds averaged, and u' is interpreted as u'$_{rms}$. The mean pressure term remains, as the coordinate transformation does not affect the magnitude of the pressure force. Thus, in the current Lagrangian formalism only the mean and fluctuating velocity in the x-directions (U and u'$^2$) are needed to determine explicitly the Reynolds stress gradient, which can then be integrated in the y-direction to find the Reynolds stress itself.

In our previous work, the DNS data were limited to Re$_\tau$ = 650 [6, 7], leading to some skepticism of the entire analysis framework. We include here the application of Eq. 1 to a higher Reynolds number from the widely-accepted data set from the Johns Hopkins University turbulence data base [16]. Figure 2 shows the comparison between the DNS data [16] and the theoretical result obtained from Eq. 1, for turbulent channel flow at Re$_\tau$ = 1000 and 5200. At these Reynolds numbers, and possibly due to the density difference, the pressure fluctuation term needs to be included, using the same transform $\frac{\partial}{\partial x} = C_1 U \frac{\partial}{\partial y}$ as used in Eq. 1. Upon these operations, the agreement between the DNS data and the current approach is quite good in Figure 2(a). At these Reynolds numbers, the turbulence structure resembles that of flows over a flat plate, where there is a spike in the u'$^2$ profiles near the wall, and the Reynolds stress has a minimum near the wall then gradually decreases to zero toward the centerline. In spite of these differences, Eq. 1 after including the pressure fluctuation gradient term works quite well in prescribing the Reynolds stress (gradient) using a minimal set of root turbulence parameters (U, $u'^2$, and P). In addition, we can look at the Reynolds stress gradient budget, using the same DNS data [16], where the $u'^2$ transport, the pressure force term, and the viscous term (the terms on the right-hand side (RHS) of Eq. 1) add up to the Reynolds stress gradient. It is an intricate yet pure momentum balance, involving only the three terms on RHS of Eq. 1. In comparison, the Reynolds stress models [3], by expanding the terms in the Eulerian coordinate frame, involve to a rather large number of terms, all of which need to be modelled at great labor and cost with little or no physical insights about the terms.

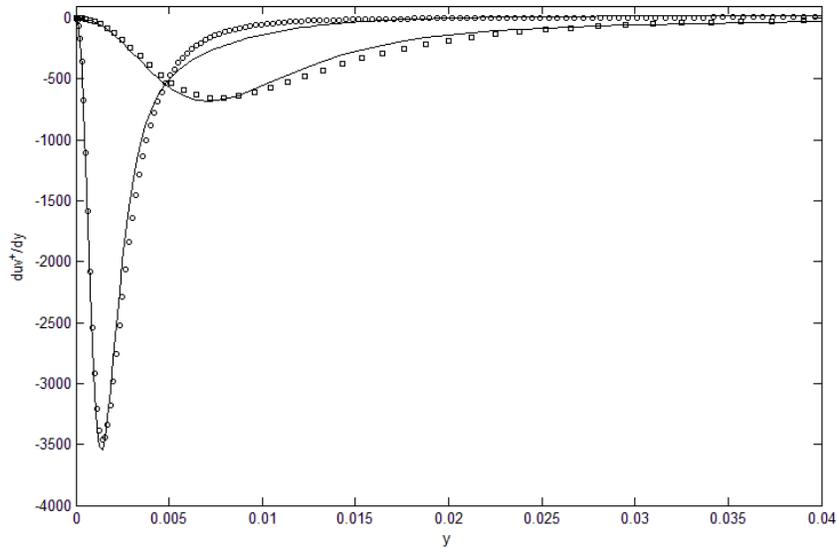

(a)

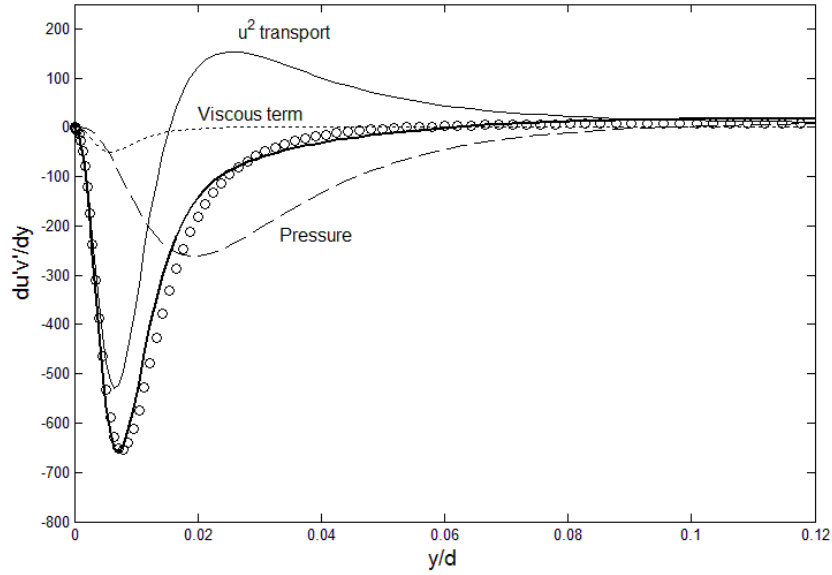

(b)

Figure 2(a). Comparison of the slope of the Reynolds stress using the current theory (Eq. 1) and DNS data [16] at $Re_\tau = 1000$ and 5200. The symbols are for the DNS data, while the lines are obtained using Eq. 1, including the pressure fluctuation term.

Figure 2(b). Reynolds stress gradient budget. DNS channel flow data (circle symbol) for $Re_\tau = 1000$ [16] are used. Solid line is the RHS side of Eq. 2, with $u^2$-transport, pressure and the viscous terms combined.

In this work, we show that using this approach the von Karman constants can be obtained that are in close agreement with published values, and that the additional equation for the Reynolds stress from the Lagrangian transport model [6, 7] renders it possible for an iterative solution method for turbulence channel flows. This is an example of the first viable solution method, based on fluid physics and minimal empiricism (if any) for an inhomogeneous turbulent flow.

**METHOD OF ANALYSIS**

In wall-bounded turbulence flows, the mean velocity profiles can be approximated as a power- or logarithm function in the so-called outer layer away from the wall. Near the wall, the viscous effects are considered dominant and the assumption of constant shear stress leads to a linear velocity profile. It is in the outer layer, where the turbulence effects are significant, that we focus our attention. In the outer layer, for sufficiently high Reynolds numbers, the viscous shear term is neglected relative to the turbulence transport, and only the momentum transport contributes to the shear stress. This shear stress in the outer layer is essentially determined by the Reynolds stress [15]:

$$dU^+/dy \sim u'^+v'^+ \qquad\qquad\qquad (2)$$

As noted in the nomenclature, the superscript "+" indicated normalization by the friction velocity.

For the logarithmic profiles, the velocity gradient is [15],

$$dU^+/dy^+ = -\kappa/y^+ \qquad\qquad\qquad (3)$$

where in the non-dimensional form of Eq. 3, $\kappa$ is the von Karman constant. The negative/positive sign in Eq. (3) is applicable if $y^+ = 0$ is chosen as the centerline/wall. Thus, the effectiveness of Eq. 1 can be tested by comparing the von Karman constant obtained from the Reynolds stress, integrated from Eq. 1; i.e. we compare u'v' with - $\kappa$/y (see Figure 4 below).

Eq. 1 for calculating the Reynolds stress, u'v' does introduce a new variable u'² variable, which is an unknown. We may use the Eulerian transport equation for u'², similar to turbulence kinetic energy equation (e.g. k in k-ε models) for this variable. Alternatively, we can derive a Lagrangian transport equation for u'², in a manner similar to the derivation of Eq. 1 [6]. This derivation leads to Eq. 4 below, with yet another higher-order term, u'²v'. However, as shown in Figure 3, this transport equation for u'² is reasonably well balanced even when we use a crude approximation, u'²v' = u'(u'v'). This additional transport equation, in principle, could be used as the third equations of turbulence transport, to solve for three unknowns, U, u'², and u'v'.

$$C_2 U \frac{\partial(<u'^3>)}{\partial y} = -\frac{\partial(<u'^2 \, v'>)}{\partial y} + 2\nu_m \left(\frac{\partial u_{rms}{'}}{\partial y}\right)^2 \qquad (4)$$

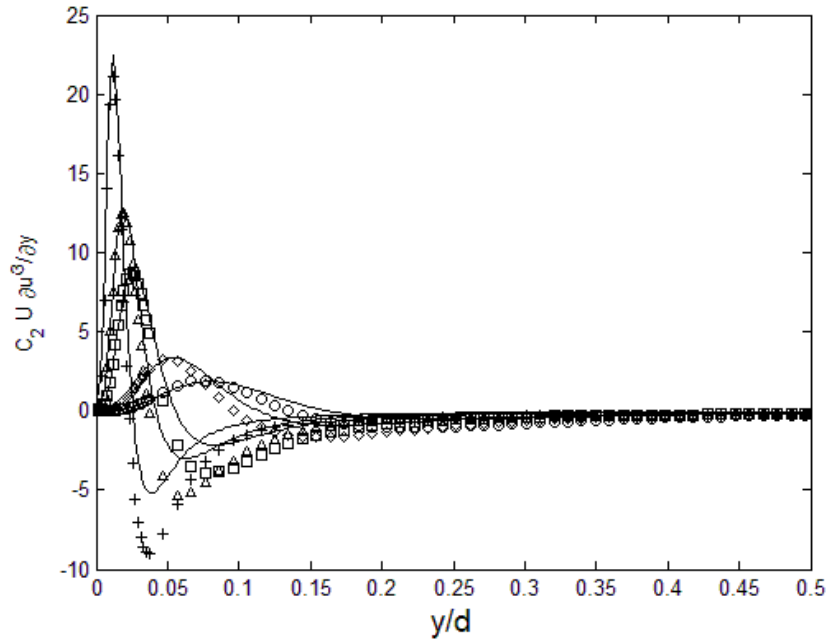

Figure 3.  Validation of Eq. 4.  Symbols are the left-hand side of Eq. 4, as computed using u'³=(u'²)³/², from DNS data [8], while lines represent the right-hand side of Eq. 4, using u'²v'=(u'²)¹/²(u'v') and u'$_{rms}$ = (u'²)¹/², also from the same DNS data.

However, in the current work, we use a simpler, iterative approach to find the von Karman constant and other key turbulence variables: First, we assume plausible functions for the mean velocity and u'² profiles.  In order to shorten the iteration, we use the power-law for the mean velocity (Eq. 5)and modified log-normal functions for the u'² profile (Eq. 6).

$$U/U_c = (y/d)^{1/n} \qquad\qquad\qquad\qquad (5)$$

$$u'^2 = A[(y/d)\exp\{-(\ln(x)-M)/S^2)/Sx\} + \delta] \qquad\qquad (6)$$

Here, n is the exponent for the velocity profile, A a pre-exponential factor, M the mean of the log-normal function, S the variance, and δ an offset (u'² does not go to zero as y/d → 1).  Using these two assumed forms, we can find an an initial estimate for the Reynolds stress with Eq. 1.  Then, the resulting velocity gradient is compared with the Reynolds stress, using Eq. 2 in the outer layer. Both the power law exponent and the modified log-normal function parameters are iterated, until convergence between the velocity gradient and the Reynolds stress is achieved.  This iterative procedure is illustrated in Fig. 4 and 6.

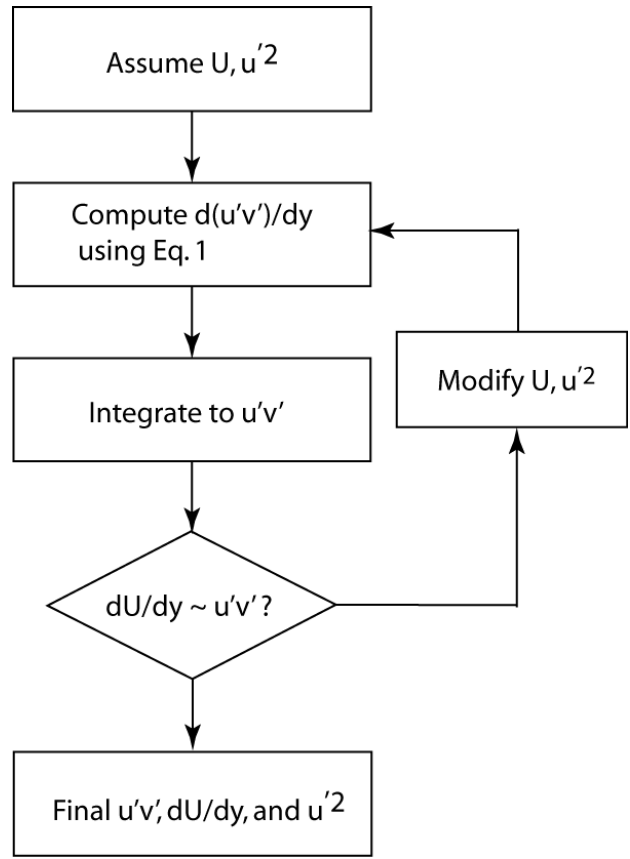

Figure 4. A flow chart for the iterative solution method to find the Reynolds stress and the mean velocity gradient (von Karman constant), described above.

**RESULTS AND DISCUSSION**

First, we check the ability of Eq. 1 to produce some key turbulence parameters. In Figure 5, the Reynolds stress profiles, obtained by integrating Eq. 1, is compared with DNS data [8]. In the outer layer, the Reynolds stress is approximately equal in magnitude to the velocity gradient (Eq. 2). If we compare Eq. 3 to the Reynolds stress profiles, then we can obtain the von Karman constants. For $Re_\tau = 300$ to 650, the logarithmic law or its gradient, $-\kappa/y$, agrees with the Reynolds stress profile in the outer layer ($y/d > 0.1$). In addition $-\kappa/y$ obviously does not go to zero at the centerline, violating the boundary condition there. Thus, the logarithmic law is an approximation.

We can recover from Eq. 1 von Karman constants that range from 0.385 to 0.444, comparable to accepted value of 0.41.

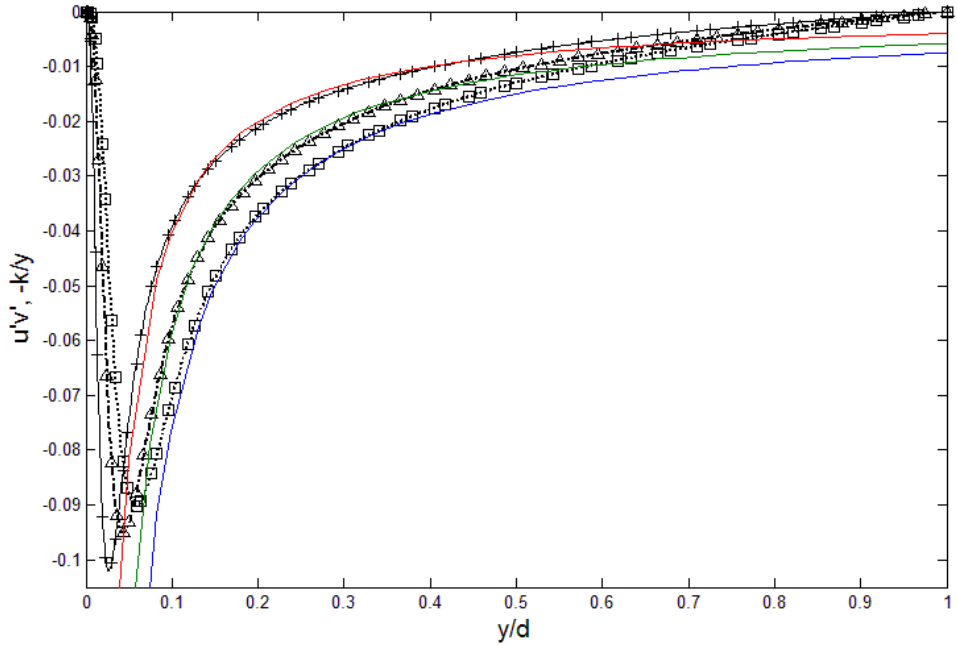

Figure 5. Comparison of the gradient of the logarithmic law and the Reynolds stress from Eq. 1. Re$_\tau$ = 300 (squares), 450 (triangles) and 650 (+). Data symbols are from DNS results [8], while lines are from Eq. 1 (similar to Figure 1e).

The accepted picture of the wall-bound turbulent flows is that the turbulence transport is dominant in the so-called "outer layer", while the flow transitions to laminar behavior close to the wall in the "inner layer". While this inner layer thickness has been estimated at 0.1 of the channel width based on observational evidence, the Reynolds stress profile can be used as an indicator of this transitional behavior. In Figure 5, we can see that the maximum magnitude (negative peaks) in the Reynolds stress moves closer to the wall as the Reynolds number increases. On either side of this peak, the Reynolds stress decreases. However, on the inner side (close to the wall) the length scale of the gradient becomes small making the laminar shear stress itself large, thus leading to the near laminar behavior. We can use the location of these negative peaks in the Reynolds stress as a transition point from turbulence- or laminar-dominated momentum transport. From Eq. 1, and its integrated Reynolds stress, this location can easily be obtained. Although limited to low Reynolds numbers ($Re_\tau < 650$), current estimate for the peak location follows, $\delta_T/d*Re_\tau^{1/2} = 19$, where $\delta_T$ is the transition point, and d the channel half-width.

Using the algorithm in Figure 4, we can use the RANS and Eq. 1 to solve for this turbulent flow, iteratively. We start by assuming U, and $u'^2$ profiles (Eqs. 5 and 6), and iterate until convergence criterion (Eq. 3) is achieved within a set error limit. This convergence process is shown in Figure 6, where the variance, S (the largest effect in on the Reynolds stress) in Eq. 6, is varied until a reasonable convergence is attained.

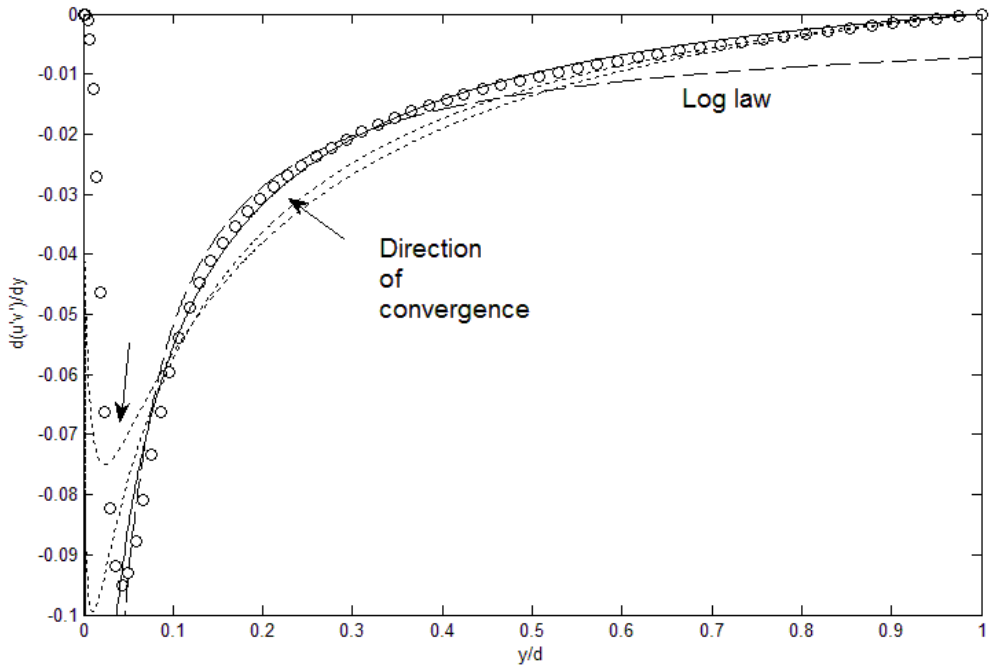

Figure 6.  Approach to the convergence condition by varying the variance (S) in the log-normal function for $u'^2$ (Eq. 6).  Dotted lines are converging solutions, while the solid line is the converged solution.  The data symbol is from the DNS data [8], for comparison.  The slope of the log law is also plotted.

Figure 7 shows the converged result, and comparison to DNS data [8] and other key parameters.  We can see that the convergence criterion is met only in the region of high positive slope, and the solution continues decrease near the wall.  However, the agreement with DNS data for the Reynolds stress, the outer solution, is quite good.  For the inner layer, the convergence criterion of Eq. 3 is not valid due to significant effect of viscous shear, so at this point no solution can be presented.  The inner Re, plotted in Figure 7, is merely a best-fit to the DNS data [8], by varying the parameters in the assumed $u'^2$ profiles.  Comparison with with $-\kappa/y$ yields a von Karman constant of 0.43 in the outer layer.

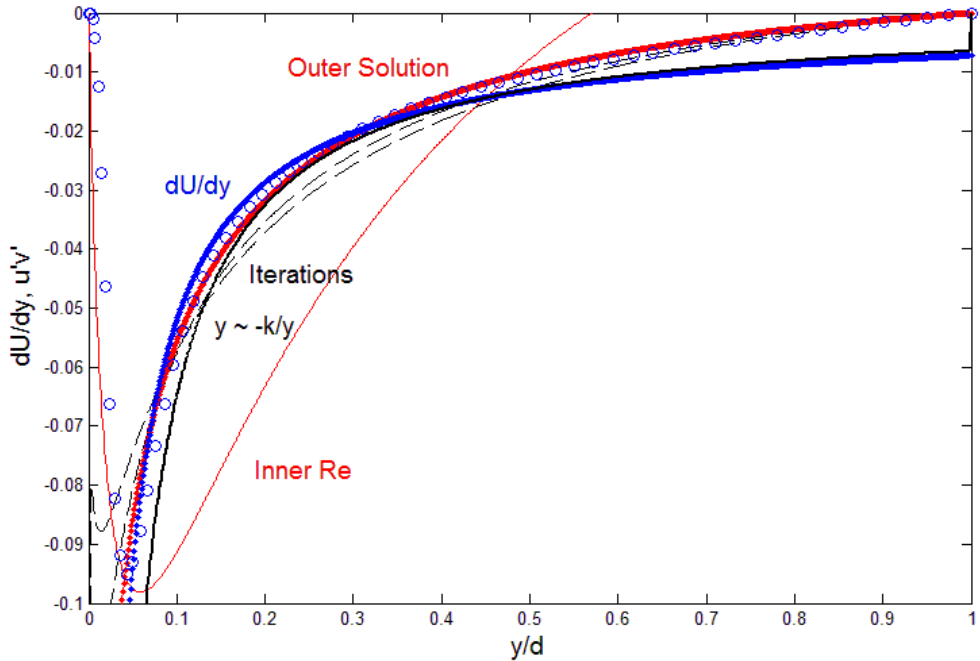

Figure 7. Converged solution for the outer layer and comparison to DNS [8] and other key turbulence parameters. $Re_\tau = 450$.

Upon convergence, we can examine the final $u'^2$ profiles, and again compare with DNS data [8] in Figure 8. The inner layer profile is simply a best-fit case, and not presented as a solution. The profiles have been reversed in the y/d axis (left for the inner layer), for easy viewing. For the outer layer, the agreement with DNS data is inexact but reasonable. Only the correct slope in the outer layer is sought in the solution procedure, and in that sense the algorithm is forgiving of the inexact iteration for the $u'^2$ profiles.

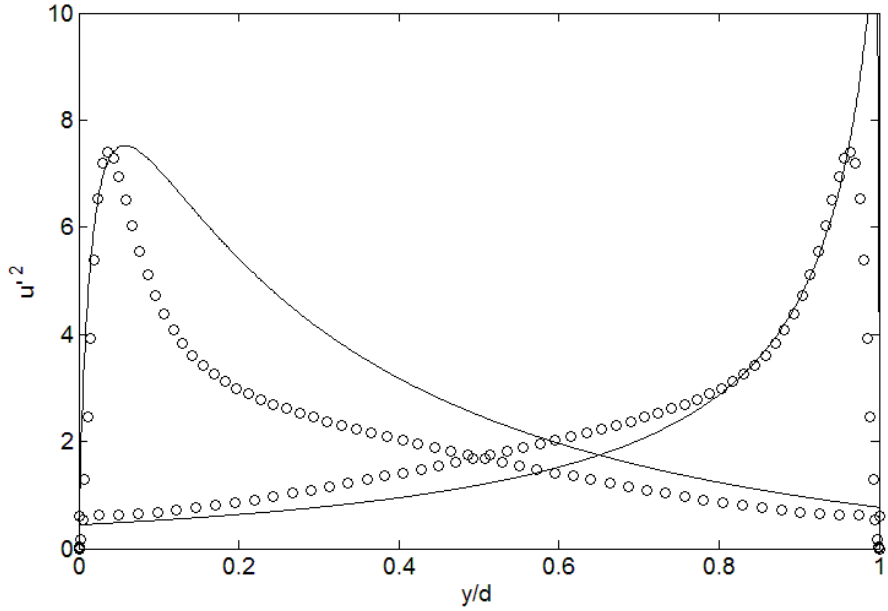

Figure 8. Final $u'^2$ profiles in the outer layer, upon convergence. $Re_\tau = 450$.

For jet flow, it is simple to check the accuracy of Eqs. 1 and 4, as the pressure term is not present in free jets, and also the viscous dissipation is quite small. Here, we again interpret $u'$ as $u'_{rms}$, and approximate $u'^2 v'$ as $u'_{rms}(u'v')$. A comparison with experimental data of Gutmark and Wygnanski [17] is shown in Fig. 9, where both $u'^3$ (Eq. 4) and $u'v'$ (Eq. 1) gradients track the data reasonably accurately. Any discrepancies are due to the above approximations, and also in extracting gradients from experimental data.

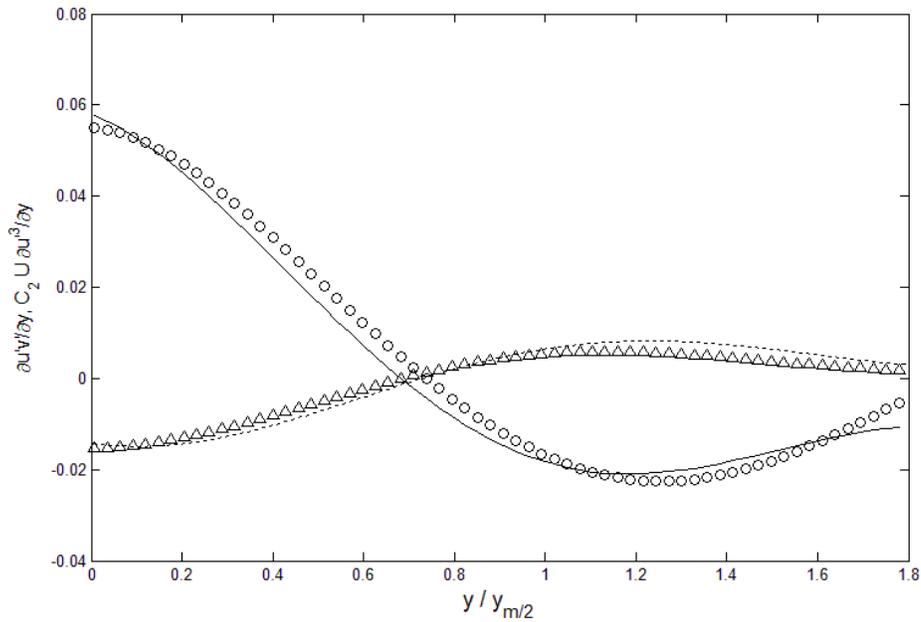

Figure 9. Validation of the transport of u'v' (Eq. 1) and u'² (Eq. 4) for turbulent jet flow. The data of Gutmark and Wygnanski [17] have been used for comparison.

As for the transport of mean momentum, it is known that the mean momentum is mainly determined by the Reynolds stress in jet flows for sufficiently high Reynolds numbers, i.e. the shear stress is dictated by the Reynolds stress [15]. Thus, we can again use the abbreviated form of RANS:

$$\mu \frac{dU}{dy} = -\rho(u'v') \qquad (7)$$

In this way, Eqs. 1, 4 and 7 furnish us with 3 equations to solve for three unknowns, U, u'², and u'v'. These equations can be solved numerically, or we can use a test function for one of the variables, u'², to prove the robustness of the equation set, the results of which are shown in Figure

10. The sequence is we input u'² test function and some initial velocity profile, U, in Eq. 1, and run them through Eq. 4 and 7, until no further changes in U or u'² profiles occur. Comparisons are again made with data by Gutmark and Wygnamski [17].

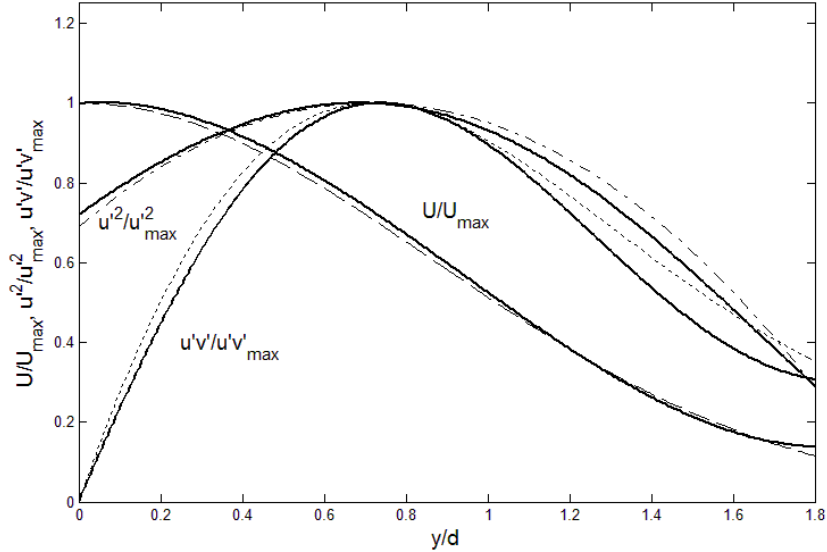

Figure 10. Mean velocity, u'², and the Reynolds stress (u'v'), all normalized by the peak value, that are obtainable from Eqs. 1, 4 1nd 7. An inverted parabola is used as a test function for u'², to initiate the integration, and then results are iterated using Eq. 2-4. Solid lines are the data of Gutmark and Wygnanski [17].

**CONCLUSIONS**

We have used the Lagrangian transport of momentum [6] to compute the Reynolds stress in terms of basic turbulence parameters. Key parameters such as the von Karman constant and inner layer thickness are obtained using this method, in reasonable agreement with accepted values. Also, the approach is verified for higher Reynolds numbers (Re$_\tau$ = 1000 and 5200) using the Johns Hopkins University turbulence data [16], where it is shown that the pressure fluctuation term must be included in the calculations of Eq. 1 and that the viscosity term is not as important at these conditions (high $Re_\tau$ and/or low density). The current approach can be combined with RANS equation, for an iterative solution in the outer layer of turbulent channel and jet flows. These are examples of viable solution method for inhomogeneous turbulent flows. Prior work [6, 7] shows that the same method is applicable in other canonical flows: boundary layer and jet flows, and it forms a basis for physics-based modeling method for turbulent flows in complex geometries.